\documentclass[aps,twocolumn,superscriptaddress,floatfix,nofootinbib,prl,longbibliography]{revtex4-1}
\usepackage[utf8]{inputenc}
\usepackage{lipsum}
\usepackage{graphicx}
\usepackage{amsmath,amssymb}
\usepackage{braket}
\usepackage{mathtools}
\usepackage{hyperref}
\usepackage{multirow}
\usepackage{color}
\usepackage[normalem]{ulem}
\usepackage{amsfonts}
\usepackage{bbm}

\renewcommand{\(}{\left(}
\renewcommand{\)}{\right)}
\newcommand{\bfvec}[1]{\mathbf{#1}}

\begin{document}

\title{A note on the quasiperiodic many-body localization transition in dimension $d>1$}

\author{Utkarsh Agrawal}
\affiliation{Department of Physics, University of Massachusetts, Amherst, MA 01003, USA}

\author{Romain Vasseur}
\affiliation{Department of Physics, University of Massachusetts, Amherst, MA 01003, USA}

\author{Sarang Gopalakrishnan}
\affiliation{Department of Physics, The Pennsylvania State University, University Park, PA 16802, USA}

\begin{abstract}

The nature of the many-body localization (MBL) transition and even the existence of the MBL phase in random many-body quantum systems have been actively debated in recent years. In spatial dimension $d>1$, there is some consensus that the MBL phase is unstable to rare thermal inclusions that can lead to an avalanche that thermalizes the whole system. In this note, we explore the possibility of MBL in quasiperiodic systems in dimension $d>1$. 
We argue that (i) the MBL phase is stable at strong enough quasiperiodic modulations for $d = 2$, and (ii) the possibility of an avalanche strongly constrains the finite-size scaling behavior of the MBL transition. We present a suggestive construction that MBL is unstable for $d \geq  3$.

\end{abstract}
\maketitle

\paragraph*{\textbf{Introduction.}} Generic isolated many-body systems are expected to thermalize under their intrinsic unitary dynamics: at long times, \emph{local} observables and correlation functions exhibit equilibrium behavior~\cite{DeutschETH, SrednickiETH, Rigol07, Huse-rev}. As already suggested in Anderson's seminal paper~\cite{Anderson58}, thermalization is not fully generic: systems subject to strong quenched randomness or quasiperiodic modulations can instead exhibit a many-body localized (MBL) phase, in which thermalization fails, and the system instead retains a local memory of its initial conditions to arbitrarily late times~\cite{Basko06, Mirlin05, BaskoMBLshort, OganesyanHuse, Huse-rev, AbaninRMP, imbrie2016many}. This memory is due to the existence in the MBL phase of (quasi-)local operators that are \emph{exact} integrals of motion~\cite{Serbyn13-1, Huse13, ros2015integrals, imbrie2016many, imbrie2017local}.
Despite the existence of a mathematical proof, under minimal assumptions~\cite{imbrie2016many,ImbriePRL}, of the existence of an MBL phase in 1d random quantum systems with short-range interactions, the existence and nature of the MBL phase as observed in numerics and in experiments remain the subject of active debate~\cite{Vidmar2019,sels2020dynamical,abanin2019distinguishing,panda2020can,crowley2020constructive,morningstar2021avalanches,PhysRevB.104.184203,leblond2020universality,2021arXiv210509348S}. Asymptotically, the MBL transition in 1d (assuming it exists) is believed to be driven by ``avalanches'' caused by rare thermal inclusions~\cite{DeRoeck2017}. This leads to a Kosterlitz-Thouless-like picture of the asymptotic transition~\cite{GoremykinaPRL,MBLKT,MorningstarHuse,PhysRevB.102.125134,vsuntajs2020ergodicity,PhysRevResearch.2.042033}, while the finite-size MBL crossover observed in numerics is driven by many-body resonances~\cite{crowley2020constructive,morningstar2021avalanches,PhysRevB.104.184203}.  Assuming the existence of the avalanche mechanism, the MBL phase is always unstable to thermalization in spatial dimensions $d>1$~\cite{DeRoeck2017, Potirniche2019}.

Many experiments on the MBL phase treat systems subject to \emph{quasiperiodic} (QP) rather than random potentials~\cite{Schreiber15, bordia2017probing, IyerQP, KhemaniCPQP, PhysRevB.96.104205, zhang2018universal, PhysRevB.98.224205,vznidarivc2018interaction, Doggen2019, doggen2021many, popperl2021dynamics,PhysRevB.103.L220201,PhysRevB.104.214201}. 
Noninteracting QP systems exhibit Anderson localization, and MBL appears at least perturbatively stable for weak enough interactions~\cite{IyerQP}.
For QP systems, unlike random systems, the avalanche instability does not rule out the possibility of an MBL phase in $d > 1$. In the random case, the existence of rare locally thermal regions is inevitable in a large enough sample; in QP systems, however, there is the possibility that the avalanche \emph{does not get started} because there are strictly no thermal regions in the sample. 
On the other hand, very little is known with certainty about MBL in QP systems: even proving that it exists in 1d has yet not been possible, because of the lack of relevant analytical tools. 


In this note, we study the nature of avalanche instabilities in QP systems, and argue that the MBL phase appears to be stable at strong enough QP modulations (though we cannot rule out other instabilities that have yet to be uncovered). We also argue that the avalanche picture strongly constrains the nature of the MBL transition in spatial dimensions $d=2$, as it reduces it to a nucleation problem. This mechanism puts strong constraints on the finite size scaling near such transitions, which we derive on general grounds. 


\paragraph*{\textbf{Setup.}} We consider a two-dimensional interacting spin-$\frac{1}{2}$ system (or alternatively, a system of interacting spinless fermions) defined on a square lattice.
 The exact details of the microscopic model will not be important for our purposes, and we will denote the strength of the interactions given by $J$. 
  A quasiperiodic (QP) field, $h_{xy}$, is applied to the system \begin{align}
h_{xy} = W_1 \cos\(2\pi \varphi\bfvec{ k_1.r} + 2\pi\phi_1 \) + W_2 \cos\(2\pi\varphi\bfvec{k_2.r} + \phi_2 \),
\end{align}
where $\bfvec{r}=\(x,y\)$, $\bfvec{k_{1,2}}$ are orthonormal vectors, $(\phi_1,\phi_2)$ are phase shifts, and $\varphi$ is the QP frequency which we take to be the golden ratio for concreteness. 
 $W_{1,2}$ determines the strength of the QP potential and for simplicity we assume $W_1=W_2=W$. 
We assume $\bfvec{k_{1,2}}$ are generic vectors and not fine tuned; for certain special values of $\bfvec{k_{1,2}}$ there are extensive lines of resonances in the system which destroy MBL for any strength $W$~\cite{Szabo2020}. We will denote by $L$ the linear size of the system. For a given QP system, we are interested in localization properties averaged over $\phi_{1,2}$. Different $\phi_{1,2}$ are equivalent to (approximate) translations over the lattice.
The MBL to thermal transition, if it exists, is tuned by the ratio of $W$ and interaction strength $J$. For large $g=W/J$ we expect a MBL phase, whereas for small $g$ the system should thermalize under its own dynamics. 

\paragraph*{\textbf{Avalanche instability in $d>1$.}} 
We first briefly review how an avalanche spreads in $d > 1$ starting from an initial spherical thermal region (``bubble'') of radius $\ell$ embedded in a putatively MBL bulk~\cite{DeRoeck2017}. (As we will discuss below this is not the dominant thermal region, but it is the simplest to analyze.) This spreading process is essentially the same in random and QP systems. We first imagine cutting all the bonds connecting the thermal region to the bulk. Then by assumption the bulk has a complete set of exponentially localized integrals of motion (LIOMs), with localization length $\xi$, while the bubble has a random-matrix like level structure with level spacing $2^{-\ell^d}$. We now recouple the bubble to the bulk. LIOMs a distance $L$ from the center of the bubble couple to the bubble with matrix element $\exp(-|L-\ell|/\xi)$. Provided that the associated Golden-Rule decay rate is larger than the level spacing of the bubble, these LIOMs get absorbed into the bubble; crucially, each time the bubble absorbs a LIOM, its level spacing goes down by a factor of two. If the bubble is initially large enough, it is sure to absorb nearby shells of LIOMs, and thus decrease its level spacing. After the bubble has grown to a size $x$, its level spacing is $2^{-x^d}$; meanwhile, the coupling to the next shell of LIOMs is $\exp(-|x-\ell|/\xi)$. Asymptotically, the level spacing decreases faster than the matrix element, so once the bubble has started growing it continues to grow. (In one dimension, by contrast, the two quantities scale the same way, allowing the MBL phase to remain stable for small enough $\xi$.)

At each stage of this procedure, the bubble grows by an amount proportional to its surface area. Since bubbles with $N$ spins are exponentially rare in $N$, the most dangerous bubbles in any dimension (i.e., the ones likeliest to start an avalanche) are those that maximize the surface area to volume ratio, i.e., one-dimensional bubbles. The level spacing of a one-dimensional bubble goes down to $2^{-(z-1)N}$ after it absorbs one layer of neighbors, where $z$ is the typical coordination number; however, its coupling to the next layer of spins is only down by a constant factor $\exp(-1/\xi)$. Thus, naively, we might estimate that $N_c \sim 1/\xi$. In fact this is an underestimate of the critical bubble size since it neglects the fact that at large $W/J$ the bubble becomes an increasingly ineffective bath on account of its narrow bandwidth~\cite{Rahul14-2}; a more careful estimate yields $N_c \sim 1/\xi^2$ in all $d > 1$~\cite{Gopalakrishnan2019}. 
Bubbles with fewer than $N_c$ spins are ineffective as baths, and absorb at most O(1) spins before they stop growing.
In a random system the probability of finding two spins to be in resonance scales as $J/W \sim 1/|\log(\xi)|$; thus the density of super-critical bubbles scales as $P(N_c) \sim \exp(-1/\xi^3)$. This is nonzero for any $\xi > 0$ in the thermodynamic limit, so a random system always thermalizes. 

For a quasiperiodic system, the same estimate of $N_c$ applies; however, since the potential is deterministic, the density of super-critical bubbles need not scale in the same way with $\xi$: in fact, it is entirely possible for this density to remain strictly zero in the thermodynamic limit, leaving the MBL phase stable. We now argue that this is indeed the case.

\begin{figure}[t!]
\centering
\vspace{-1cm}
\includegraphics[width=0.5\textwidth]{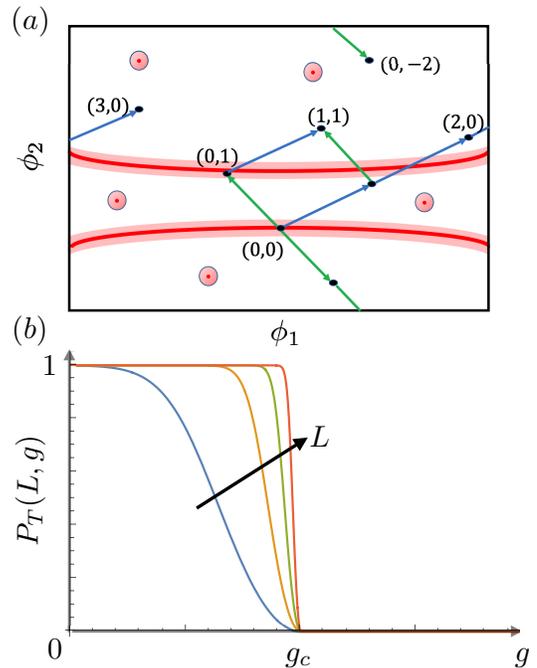}
\vspace{-1.5cm}

\caption{(a) Lattice points and $\delta$-resonances on $\phi_1-\phi_2$ torus. The plot is a schematic and is not drawn to scales. The lattice points are defined as $\phi_1(\bfvec{r})=\varphi\bfvec{k_1.r}+\phi_{0,1}$ and $\phi_2(\bfvec{r})=\varphi\bfvec{k_2.r}+\phi_{0,2}$. Bold red points and lines correspond to locations of perfect 3 and 2 site resonances respectively; these bold regions do not form a dense set. Light red regions around the bold red points and lines correspond to regions where interactions will hybridize 3 and 2 sites to form a resonance; the radius of the light red region $\delta_\epsilon$ increases with the interaction strength $J=\epsilon$ (see main text). To form bigger resonances, nearby lattice points should lie within the red regions. In the plot above there are pairs of neighboring 2-site resonances at (0,0) and (0,1) which will be combined to a 4-site resonance by the interactions. Such 4-site resonance can be shown to always exist. Forming 5-site or higher resonances needs fine-tuning at small $\delta_\epsilon$. (b) Sketch of the finite-size scaling of the thermalization probability $P_T(L,g)$ in the nucleation scenario, see eq.~\eqref{eq: scaling ansatz}.
}\label{Fig. toy dig}
\end{figure}

\paragraph*{\textbf{Stability of the MBL phase.}} 
We consider an infinite system with weak interactions and try to identify the largest bubble that can be created by the interactions. We map the lattice points to the $\phi_1-\phi_2$ torus by identifying $\bfvec{r}$ with $\phi_1(\bfvec{r})=\varphi\bfvec{k_1.r}+\phi^0_{1}$ and $\phi_2(\bfvec{r})=\varphi\bfvec{k_2.r}+\phi^0_{2}$. The lattice points form a dense subset on the torus. We next identify what we call $\delta$-resonances: a connected path of $n$ sites is called $n$-site $\delta$-resonance if the difference between neighboring $h_{xy}$ along the path is within $\delta$. That is, two neighboring points $P,Q$ are part of a $\delta$-resonance if $|h_P-h_Q|<\delta$. We argue that for a fixed set of generic $\bfvec{k_{1,2}}$, and low enough $\delta$, we can only have 2- and 3-site $\delta$-resonances. To see this let us take $\delta=0$ and without loss of generality consider conditions on $\phi^0_1,\phi^0_2$ for having a $\delta$-resonance at the origin.
Demanding a two-site perfect resonance gives a constraint on $\phi^0_{1,2}$, and since we have two degrees of freedom, $\phi^0_{1,2}$, for a given value of $\phi^0_1$ (or $\phi^0_2$) there exists a $\phi^0_2$ (or $\phi^0_1$) for which there is a two-site perfect resonance at the origin. Thus these type of resonances form lines on the torus (Fig.~\ref{Fig. toy dig}). Similarly for three-site $\delta$-resonance we have two constraints which lead to isolated pairs of $(\phi_1,\phi_2)$. 
Note that the values of phases $(\phi^0_1,\phi^0_2)$ for a $\delta=0$-resonance at the origin form a measure zero subset on the torus. There is a perfect resonance at $\bfvec{r}$ if $(\phi_1(\bfvec{r}),\phi_2(\bfvec{r}))$ lies in this set. Since these points form a non-dense subset, in general there can only be at most one ${\delta=0}$-resonance in a system. 
On slightly increasing $\delta>0$, the points on the torus consistent with $\delta=0$-resonances become balls of small radius proportional to $\delta$; see Fig.~\ref{Fig. toy dig}. There is $\delta$-resonance at $\bfvec{r}$ if $(\phi_1(\bfvec{r}),\phi_2(\bfvec{r}))$ lies in one of these balls. Since lattice points are dense on the torus, there are non-zero densities of 2- and 3-site $\delta$-resonances which go up with $\delta$. But for a higher site $\delta$-resonance to form we require \textit{overlapping} 2-sites and 3-sites resonances which is generically not possible without fine-tuning $\bfvec{k}$.

\begin{figure}[t!]
\centering
\includegraphics[width=0.5\textwidth]{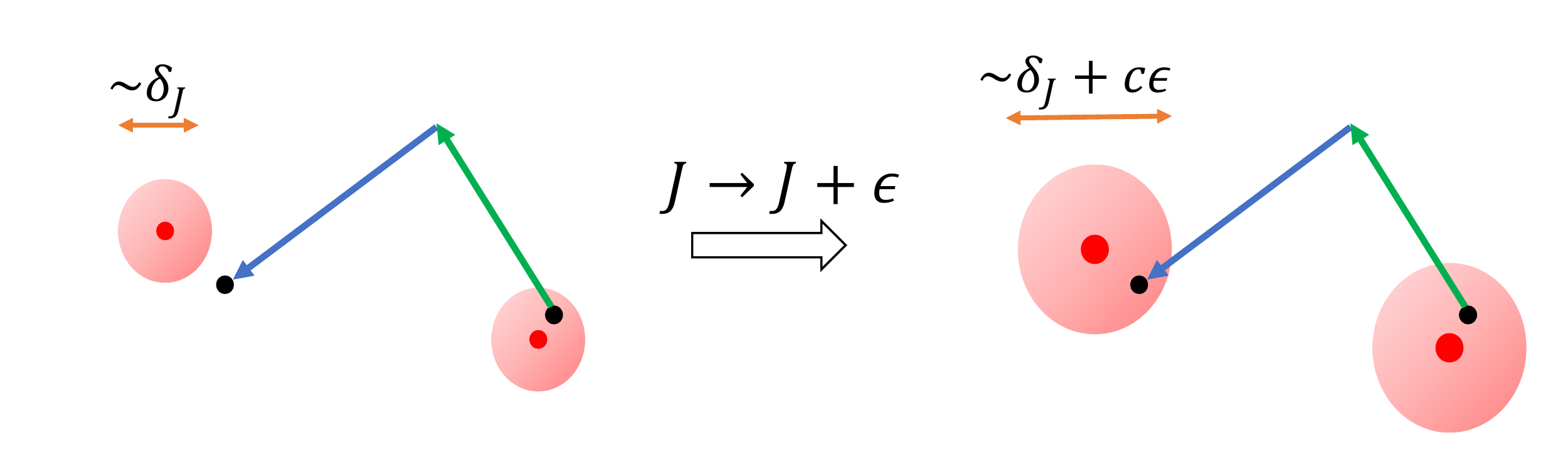}
\caption{Here we demonstrate a hypothetical scenario where a small $\epsilon$ increase in the interaction strength $J$ leads to formation of two 3-site resonances separated by 2 lattice sites.}\label{Fig. toy dig2}
\end{figure}

Interactions will hybridize the $\delta$-resonant sites. For small interaction strength $J=\epsilon$ all $\delta<\delta_\epsilon$ resonances will be hybridized with $\delta_\epsilon$ depending on $\epsilon$ and $\delta_\epsilon\rightarrow 0$ as $\epsilon\rightarrow 0$; we will refer to the $\delta$-resonances hybridized by the interactions simply as resonances. 
The exact relation between $\delta_\epsilon$ and $\epsilon$ is not important. Moreover, interactions will also cause nearby resonances to hybridize~\cite{Gopalakrishnan2019}. For example, if two 3-site resonances are, let's say, 2 lattice sites\footnote{The distance on which the interactions are able to merge two resonances scale as $-\frac{1}{\ln J/W}$ which for small $J$ is a small number. This can also be thought of as the localization length.} 
away then there is a chance that interactions will hybridize them to form a 6-site resonance; see Fig.~\ref{Fig. toy dig2}. We can show that there always exists a pair of neighboring 2-site $\delta$-resonances which will be hybridized to a 4-site resonance by the interactions. To get even bigger resonances we need the 2,3,4-site resonances to get close to each other. Again for small enough $\epsilon$, we expect this to require fine-tuning of $\varphi,\bfvec{k_{1,2}}$, so we conjecture that there are no 5-site or higher resonances for weak enough interactions. 

Note that 2,3,4-site resonances are not strong enough thermal bubbles to thermalize nearby single spins. Thus the only way we can get bigger resonances/baths is by having multi-site resonances close together which, as mentioned above, requires fine-tuning of $\bfvec{k_{1,2}}$ due to the correlated nature of the QP potential.
The absence of bubbles of size greater than $5$ for low enough $J$ implies no avalanche instability to the MBL as the critical bath size, which scales as $\sim1/\xi^2$, can be made arbitrarily large by choosing lower $J$ or higher $W$. Thus we argue that the MBL phase is stable for large enough $g=W/J$. 

We remark, finally, that it is possible to choose quasiperiodic potentials for which even two-site resonances are absent~\cite{PhysRevLett.53.1582}. Thus the size of the largest bubble is nonuniversal and depends on the details of the quasiperiodic potential; what is generic, however, is that the size is $O(1)$, and therefore is inadequate to create a super-critical bubble when $g$ is large enough.

\paragraph*{\textbf{Constraints on the MBL transition.}} Having argued for the stability of the MBL phase for strong enough QP potentials $g \gg 1$, we now discuss possible scenarios for its transition to the thermal phase. As $g$ is decreased, the criterion for two sites (or pairs of sites, etc.) to be resonant becomes less stringent so the density of resonances increases, and some of them overlap forming larger resonances [Fig.~\ref{Fig. toy dig2}]. Eventually these resonances will form a bath and thermalize the whole system. We discuss one natural scenarios by which this might happen: the \emph{nucleation} scenario. If this scenario holds we can strongly constrain the properties of the transition. Of course, the true transition might take place through an entirely different instability, and if it does so our constraints will not apply.

We first comment on the nucleation scenario. The QP MBL phase in $d > 1$ is reminiscent of a supercooled liquid: it is unstable to the introduction of any thermal bubbles, but in the absence of thermal bubbles the instability cannot get started. In the nucleation scenario, the transition happens because a local configuration somewhere in the system is able to get large and well-coupled enough to act as a heat bath and seed the avalanche. The key assumption is that when this happens the seed is of finite extent, i.e., there is some $L$ such that regions of linear size $\geq L$ have some probability $P_T(L)$ of thermalizing. In this scenario the seed is a discrete but ``good'' (i.e., finite-bandwidth) bath at the point when it first comes into existence. If $P_T(L) > 0$ for some $L$, then a large enough system will necessarily thermalize, so in the MBL phase we require that $P_T(L) = 0$ for all $L$.

Moreover, in the thermal phase, near the critical point, we expect $P_T(L,g)$ to scale as $\sim \rho_c(g)L^2$ where $\rho_c(g)$ is the density of the critical thermal bubbles. Due to the deterministic and almost periodic nature of the QP potential, the critical bubbles appear in a QP pattern with a characteristic length scale $\zeta$~\cite{Agrawal2019,Agrawal2020} which is also equal to the correlation length, $\(g_c-g\)^{-\nu}$, inside the thermal phase. Thus, $\rho_c\sim \zeta^{-2} \sim\(g_c-g\)^{2\nu}$.
Based on the discussion above, we propose following scaling ansatz for $P_T(L,g)$,\begin{align}
P_T(L,g)&=f\( (g-g_c)L^{\frac{1}{\nu}} \) \nonumber \\
&= \begin{cases}
\sim 1 & g<g_c, \ L\gg\(g_c-g\)^{-\nu} \\
\sim L^2 \(g_c-g\)^{2\nu} &  g<g_c, \ L\ll \(g_c-g\)^{-\nu} \\
0 & g>g_c
\end{cases}. \label{eq: scaling ansatz}
\end{align}
Note that this particular finite size scaling form forbids the existence of a finite size crossing: $P_T(L,g)$ must be an increasing function of $L$. The exponent $\nu \geq 1/2$ is constrained by the Harris-Luck criterion~\cite{Luck_1993}. 

The nucleation scenario assumes that thermalization is due to a finite-size resonant configuration that becomes \emph{possible} when $g$ decreases to some critical value. This is not the only possible scenario: a possible alternative is that as one comes from the thermal phase, increasing $g$, the nucleus of the avalanche becomes increasingly sparse in real space. The spacing of sites on this sparse resonant cluster might go as $\ell \sim |g - g_c|^{-\gamma}$. As it does so, its bandwidth shrinks as $\delta \sim \exp(-\ell/\ell_0)$, and it becomes an increasingly ineffective bath. In this scenario, a thermal seed will repeat at a distance scale where the quasiperiodic potential repeats to exponential accuracy. This would give Kosterlitz-Thouless-like scaling of the form $\zeta \sim \exp(|g - g_c|^{-\gamma})$.

\paragraph*{\textbf{Weak randomness and Hartree effects}.} We now briefly discuss the fate of the QP MBL phase when a weak random potential of bandwidth $w_r$ is added to the quasiperiodic potential. When the random potential is sufficiently weak, it does not affect the counting of few-spin resonances that we gave above. However, at distances $\sim 1/\sqrt{w_r}$, one can typically find pairs of sites that are detuned by less than $w_r$. Rare fluctuations of the random potential can then put sequences of sites at this distance scale exactly on shell, creating a sparse thermalizing network of very narrow bandwidth~\cite{Rahul14-2} that can initiate an avalanche.

This observation raises a natural concern about our analysis of resonances: while checking for resonance, we ignored the random ``Hartree'' shifts of the splitting of a LIOM due to its interactions with other LIOMs that (at high temperature) are in a random configuration. Could these Hartree shifts act like emergent randomness and create a resonant network? 
The answer to this question, as we will now discuss, differs between $d = 2$ and $d > 2$. 

We first consider $d = 3$ (though our considerations extend naturally to $d > 3$). Let us consider two sites $A = (0,0,0)$ and $B = (L,0,0)$ that are detuned by an amount $\leq J$ before including Hartree effects. Here we expect $L \sim J^{-1/3}$. We now attempt to bring these sites closer to resonance by adjusting the occupation pattern of LIOMs on the planes $(0,x,y)$ and $(L,x,y)$. The result of Ref.~\cite{Rahul14} shows that one can get the two energies arbitrarily close to one another.  This is a consequence of the fact that the thermally averaged spectral function of a LIOM in $d \geq 2$ forms a continuum. Moreover, for large $L$ (i.e., small $J$), the two planes are far apart so ``crosstalk'' between the two planes is negligible. This procedure clearly allows us to construct an arbitrarily long chain of resonances, which can then act as a thermalizing bath.
These considerations strongly suggest that in $d \geq 3$ it is impossible to construct stable LIOMS, even for QP potentials.

The direct generalization of this argument to $d = 2$ does not yield an instability, since the mechanism would involve trying to tune the LIOM energy by adjusting occupation numbers on a \emph{line}. In the limit of strong disorder, the spectral function averaged over the occupation numbers on a line is \emph{not} a continuum, but instead occupies a measure-zero set of frequencies~\cite{Rahul14}. However,  one can generalize it~\cite{DrewPrivateComm} by considering quasi-1d strips wide enough so that  the spectral function of each strip is a continuum~\cite{Rahul14}: we hope to come back to this potential instability in future works~\cite{InPrep}.

\paragraph*{\textbf{Summary}.}
In this note, we studied stability of the MBL phase in the presence of QP disorder in $d > 1$ many-body quantum systems. In 2D we find that unlike the random case, for strong enough QP strength or/and weak interactions the QP potential landscape is devoid of arbitrarily large resonant spots. 
Bigger resonances can only be formed by the combination of small resonances. However, since the set of small resonances is not dense, having multiple such resonances close to each other requires fine-tuning.
We also study the nature of transition from the MBL phase to the thermal phase using the avalanche instability. The critical size for the thermal bubble needed to thermalize the whole system does not scale with the system size. This puts  severe constraints on the finite size scaling ansatz for the transition (eq~\eqref{eq: scaling ansatz}). In $d \geq 3$, at the level of counting small resonances, the situation is very similar to $d = 2$. However, Hartree interactions among the putative LIOMs can potentially destabilize these LIOMs by setting them on resonance with one another.


Though we argue that the nature of the avalanche instability constrains the behavior of the transition, it would be interesting to verify this numerically. Finite size numerics based on exact diagonalization may not be sufficient as we need to access system sizes bigger than the critical bubble size to observe the scaling ansatz in eq~\eqref{eq: scaling ansatz}. Another possibility would be to have phenomenological renormalization group (RG) studies similar to those in 1d~\cite{AltmanRG14,Potter15,Dumitrescu17,PhysRevB.93.224201,MullerRG,PhysRevLett.121.140601,GoremykinaPRL,MorningstarHuse,PhysRevB.102.125134}. But unlike 1d, 2d systems can have non-trivial bipartitioning geometries for dividing the system into thermal and insulating regions. Nevertheless, as long as the phenomenological description of the RG contains the essential features of the avalanche instability highlighted in this paper, we should expect to see the proposed scaling ansatz near the transition. 

Another important direction for the future would be to put the arguments for stability of the MBL phase on a solid mathematical grounds. Our argument relied on the non-denseness of the resonance regions (light red regions in Fig. \ref{Fig. toy dig}) in the $\phi_1-\phi_2$ plane which suggests that nearby lattice points do not simultaneously lie in resonant regions and hence cannot be hybridized by the interactions. However we cannot rule out these instances due to some non-obvious/hidden patterns in the correlated nature of the QP potential, though they seem highly unlikely without fine-tuning of the QP potential. 

Unlike 1d, the peculiar nature of the avalanche instability enabled us to identify the universal features of the MBL transition in 2d. On the other hand, a complete understanding of 1d QP MBL transition remain missing~\cite{KhemaniCPQP} and is an obvious future direction to explore. The RG schemes have been very successful for random systems but have failed to give a consistent understanding for the QP systems. We think that combining ideas about microscopic QP resonant structures presented in this paper together with existing phenomenological RGs might help shed some light on the nature of this transition in 1d.

\paragraph*{\textbf{Note added:}} While preparing this manuscript, we became aware of Ref.~\cite{PhilAnushya} which also argues, from a complementary perspective, for the stability of 2d QP MBL phase against avalanches. After posting, we also became aware of Ref.~\cite{2022arXiv220405198S} which also argues for the stability of 2d QP MBL. 

\paragraph*{\textbf{Acknowledgments.}}  We thank D. Huse, V. Khemani, A. Morningstar, A.C. Potter, Z. Shi, H. Singh, and B. Ware for useful discussions and collaboration on related works. We acknowledge support from NSF Grants No. DMR-2103938 (S.G.), DMR-2104141 (R.V.), the Alfred P. Sloan Foundation through a Sloan Research Fellowship (R.V.).
\bibliography{mbl}
\end{document}